\documentclass[12pt]{iopart}
\usepackage{iopams}  
\usepackage{graphicx}  

\begin{document}

\title[On clustering measures for small-world networks]{Mean clustering coefficients--The role of isolated nodes and leafs on clustering measures for small-world networks}

\author{Marcus Kaiser$^{1,2*}$}
\address{$^1$ School of Computing Science, Newcastle University, Claremont Tower, Newcastle upon Tyne NE1 7RU, United Kingdom}
\address{$^2$ Institute of Neuroscience, Newcastle University, Framlington Place, Newcastle upon Tyne NE2 4HH, United Kingdom}
\ead{m.kaiser@ncl.ac.uk}

\begin{abstract}
Many networks exhibit the small-world property of the neighborhood connectivity being higher than in comparable random networks. However, the standard measure of local neighborhood clustering is typically not defined if a node has one or no neighbor. In such cases, local clustering has traditionally been set to zero and this value influenced the global clustering coefficient. Such a procedure leads to under-estimation of the neighborhood clustering in sparse networks. We propose to include $\theta$  as the proportion of leafs and isolated nodes to estimate the contribution of these cases and provide a formula for estimating a clustering coefficient excluding these cases from the Watts \& Strogatz \cite{Watts1998} definition of the clustering coefficient. Excluding leafs and isolated nodes leads to values which are up to 140\% higher than the traditional values for the observed networks indicating that neighborhood connectivity is normally underestimated. We find that the definition of the clustering coefficient has a major effect when comparing different networks. For metabolic networks of 43 organisms, relations changed for 58\% of the comparisons when a different definition was applied. We also show that the definition influences small-world features and that the classification can change from non-small-world to small-world network. We discuss the use of an alternative measure, disconnectedness $D$, which is less influenced by leafs and isolated nodes.
\end{abstract}

\pacs{89.75.Hc,89.75.Fb,87.85.mk}
\submitto{\NJP}
\maketitle

\section{Introduction}
Many real-world networks show properties of small-world networks as their neighborhood connectivity, generally denoted by the clustering coefficient, is higher than in comparable random networks\cite{Watts1998}. The local clustering coefficient for an individual node $i$ with $deg_i$ neighbors and $\Gamma_i$ edges between its neighbors is 
\begin{equation}
\label{eq.1}
C_i=\frac{\Gamma_i}{deg_i(deg_i-1)}
\end{equation}
This formula is basically not defined if the number of neighbors $deg_i$ becomes zero or one as the denominator becomes zero \cite{Costa2007b}. These cases are usually treated as $C_i = 0$ although some authors also set these values to one~\cite{Brandes2005}.  In the current scheme, these values would be part of the global calculation 
\begin{equation}
\label{eq.2}
C_1=\frac{1}{N}\sum C_i
\end{equation}
In addition, we tested an alternative and more widely used definition of the clustering coefficient~\cite{Newman2001} in which 
\begin{equation}
\label{eq.3}
C_2=\frac{\sum \Gamma_i}{\sum deg_i(deg_i-1)}
\end{equation}
This might lead to biased assessments of neighborhood clustering in the sense that values that are not defined (division by zero) should not be included in the averaging. Thus, instead of using $N$ as the number of evaluated nodes for the global $C_1$, a new number $N'$ indicating all nodes with defined local clustering should be used for a global measure $C'$.  We show that using such an adjusted measure for the clustering coefficient has several implications for network analysis and can help to identify the contribution of leafs and isolated nodes on average clustering.

On a conceptual level, the adjusted value $C'$ is more intuitive as the clustering coefficient is commonly called a measure of neighborhood connectivity: If 30\% of the local coefficients are zeros from cases where no neighbors exist, how can the classical definitions still give information about neighborhood? Cases of leafs and isolated nodes are more likely in sparse networks where the edge density $d$, the number of existing divided by the number of possible connections ($d = E/N*(N-1)$ for a network with $N$ nodes and $E$ directed edges or arcs), is low. Therefore, the classical definition is a mixed measure of neighborhood clustering and sparseness (edge density) or -- more precisely -- the frequency of leafs and isolated nodes. 

A general problem of network measures, such as the clustering coefficient, is whether sampling or perturbations change the values of these measures. Network measures are frequently used for the classification of different networks \cite{Amaral2000} or of topological changes (addition or deletion of nodes or edges) within the same network. Incomplete sampling -- only observing a sub-network of a larger network -- can lead to the wrong classification of a network as being a scale-free network \cite{Stumpf2005}. This occurred, for example, for comparing the partial and complete protein-protein interaction networks \cite{Han2005} and the router and underlying communication network \cite{Lakhina2003}. In addition to sampling, false scale-free classifications can also arise due to statistical errors \cite{Khanin2006}. Whereas previous studies investigated the effect of sampling on the degree distribution, a recent study \cite{VillasBoas2008} looked at the sensitivity to sampling and network perturbation for a range of measures: The clustering coefficient, as well as the hierarchical clustering coefficient, the hierarchical degree, and the divergence ratio were found to be least sensitive to perturbations of the topology. Therefore, classifications using the clustering coefficient (e.g. small-world classification \cite{Watts1998}) are less affected by the sampling problem. However, as we show here, the definition of the clustering coefficient can have a considerable effect on network classification.

\section{Materials and Methods}
\subsection{Networks}
We tested the effect of different definitions for the clustering coefficient on several real-world networks. All but one network, the German highway system, were small-world networks. The \textit{Caenorhabditis elegans} neuronal network consisted of individual neurons as nodes and existing synaptic connections as edges~\cite{Kaiser2006}. The metabolic networks of \textit{C. elegans}, \textit{Saccharomyces cerevisiae}, and 41 other organisms included metabolic substrates as nodes and reactions as edges~\cite{Jeong2000}. The protein-protein interaction network of \textit{S. cerevisiae} (yeast) included proteins as well as interactions as discovered by the yeast two-hybrid method (http://dip.doe-mbi.ucla.edu, dataset from 2 Dec 2007). The German highway (Autobahn) system consisted of location nodes (that is, highway exits) and road links between them (Autobahn-Informations-System, AIS, from http://www.bast.de)~\cite{Kaiser2004d}. Only the gross level of highways were included in the analysis, discarding smaller and local roads ('Bundesstrassen' and 'Landstrassen'). For the power grid, nodes represent generators, transformers and substations, and edges represent high-voltage transmission lines between them~\cite{Watts1998}. For the world-wide-web, individual pages are the nodes and links between them the edges~\cite{Huberman1999}. Information about the size of the networks as well as a reference to the source of the datasets is included in Table~\ref{tab.2}. For comparisons, we also generated random networks with the same number of nodes and edges as the original networks described above. In such Erd{\"o}s-R{\'e}nyi random networks \cite{Erdoes1960}, the probability $p$ that an individual connection between two nodes is established equals the edge density $d$ of the desired network.

\begin{table*}
\caption{Number of nodes $N$, edge density $d$, ratio $\theta$ of nodes with less than two neighbors, and factor of increase ($C_1\rightarrow C_2$) for several biological and artificial networks: \textit{C. elegans}  neuronal~\cite{Kaiser2006} and metabolic network~\cite{Jeong2000}, yeast metabolic interaction network~\cite{Jeong2001}, yeast protein-protein interaction network (http://dip.doe-mbi.ucla.edu, dataset from 2 Dec 2007), German autobahn system~\cite{Kaiser2004d}, electrical power grid of the western United States~\cite{Watts1998}, and world-wide web~\cite{Huberman1999}.}
\label{tab.2}
\begin{tabular}{lrlll}
Network & $N$ & $d$ & $\theta$ & f\\
\hline
\textit{C. elegans}$_{\mbox{\small neuronal}}$	   & 277 &	0.0275 & 0.0217 & 1.02\\ 
\textit{C. elegans}$_{\mbox{\small metabolic}}$    & 452 & 0.0106 & 0.1416 & 1.17\\ 
\textit{S. cerevisiae}$_{\mbox{\small metabolic}}$ & 551 & 0.0092 & 0.1198 & 1.14\\
\textit{S. cerevisiae}$_{\mbox{\small PPI}}$	     & 4,931	& 0.00143 & 0.2294 & 1.30\\
German highways 											 & 1,168 & 0.0018 & 0.0865 & 1.09\\
Power grid 														 & 4,677 & 0.000572 & 0.2609 & 1.35\\
World-wide web 												 & 325,729 & 0.0000138 & 0.5868 & 2.42\\
\hline
\end{tabular}
\end{table*}

\subsection{Adjusted clustering coefficient definition}
In addition to the two definitions for neighborhood clustering defined in the introduction, we looked at the effect of removing nodes with less than two neighbors corresponding to leafs and isolated nodes before averaging for the global clustering coefficient. The relation between the new coefficient $C'$ and the traditional measure $C_1$ can be derived from the fraction of nodes that have one or zero neighbors, $\theta$ by
\begin{equation}
\label{eq.4}
C'=\frac{1}{1-\theta}C_1
\end{equation}
Therefore, $$f=\frac{1}{1-\theta}$$ is the factor of the increase of the clustering coefficient $C_1$ by using the new method. Unfortunately, there is no easy transformation between the new measure $C'$ and the other measure $C_2$ (e.g. the correlation between the two measures is $r=0.06$ for 43 metabolic networks). 

\section{Results}
What is the effect of the adjusted definition $C'$ above? If one third of local coefficients were undefined, for example, the clustering coefficient would increase by 50\% and would double if half of the nodes were undefined. For the yeast protein-protein interaction network with 4,931 nodes the clustering coefficients $C_1$ and $C_2$ raised from 14.4\% and 8.4\%, respectively to 18.7\% for C'. That means that the value increased by 30\% compared to $C_1$ and more than doubled compared to $C_2$. For several real-world networks (Tab.~\ref{tab.2}), values of neighborhood connectivity increased by factors between 1.02 and 2.42; that means that the average clustering coefficient increased by up to 142\%. This indicates that current definitions significantly underestimate neighborhood clustering. For Erd{\"o}s-R{\'e}nyi random networks \cite{Erdoes1960} with the same number of nodes and edges as the yeast protein interaction network the increase was maximally 4.3\% and on average 0.7\% for 100 generated networks; thus the new clustering coefficient is still comparable with the edge density of random networks.    
 
\subsection{Network comparison}    
In addition to the effect for single networks, measures such as the clustering coefficient are often used for comparing networks. Network comparisons can either involve different original networks or the same network before and after structural perturbations. Previous studies compared the clustering coefficient of 43 metabolic networks \cite{Ravasz2002} and changes in neural correlation networks for Alzheimer \cite{Stam2007}, schizophrenia \cite{Micheloyannis2006}, and epilepsy \cite{Ponten2007,Srinivas2007} patients.  

\begin{figure}
\includegraphics{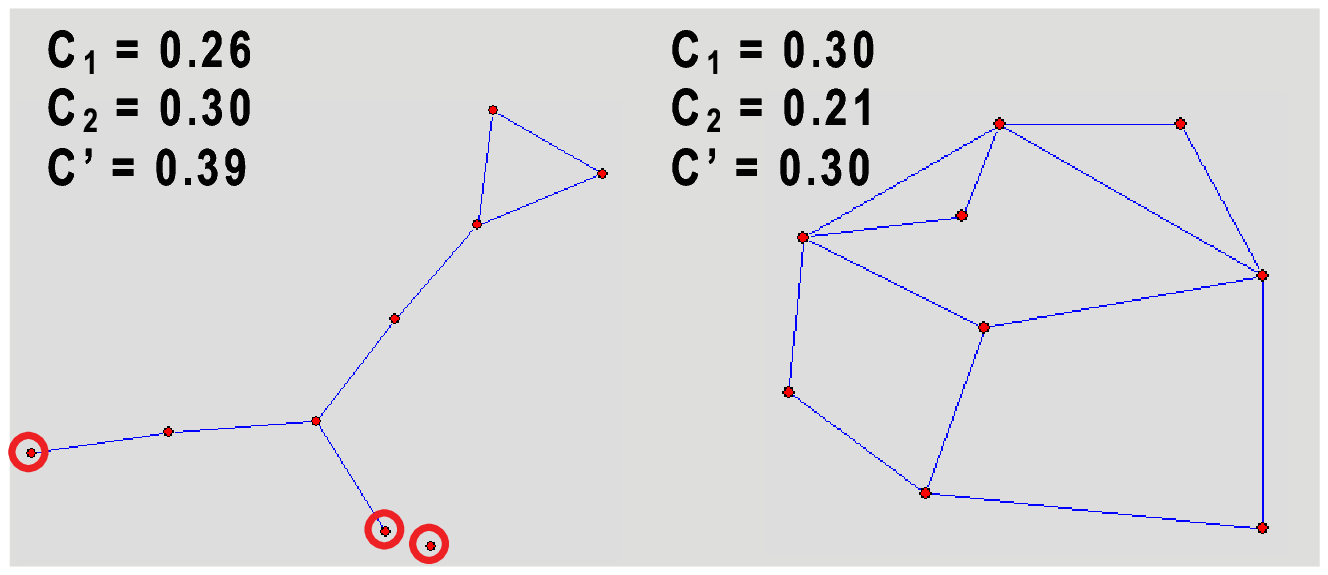}
\caption{Comparison of clustering coefficients for a sparse (left) and a dense (right) network with nine nodes. Whereas the clustering coefficient is higher in the dense network for the standard measure $C_1$, it is higher in the sparse network for the novel, $C_2$, and adjusted, $C'$, neighborhood clustering. For the adjusted clustering coefficient, isolated nodes or nodes with only one neighbor (indicated here by red circles) are excluded from the averaging.}
\label{fig.1}
\end{figure}

For network comparisons, the definition of neighborhood connectivity is critical for the comparison (Fig.~\ref{fig.1}). Assuming that we have two networks $G_a$ and $G_b$, where the first has higher classical clustering ($C_1$ measure) than the second one, i.e. $C_a>C_b$. Then, this relation will swap for the new definition to $C_a'<C_b'$ if $\frac{C_a}{C_b}<\frac{1-\theta_a}{1-\theta_b}$. Let us look at the simpler case where we compare a sparse network with a dense network still under the assumption that $C_a>C_b$. As the dense network has almost no nodes that are isolated or leafs, we can set $\theta_b=0$. Then, using the new definition $C_a'<C_b'$ if $\theta_a<1-\frac{C_a}{C_b}$. How often do these swaps occur in real-world networks?

\begin{figure}
\includegraphics{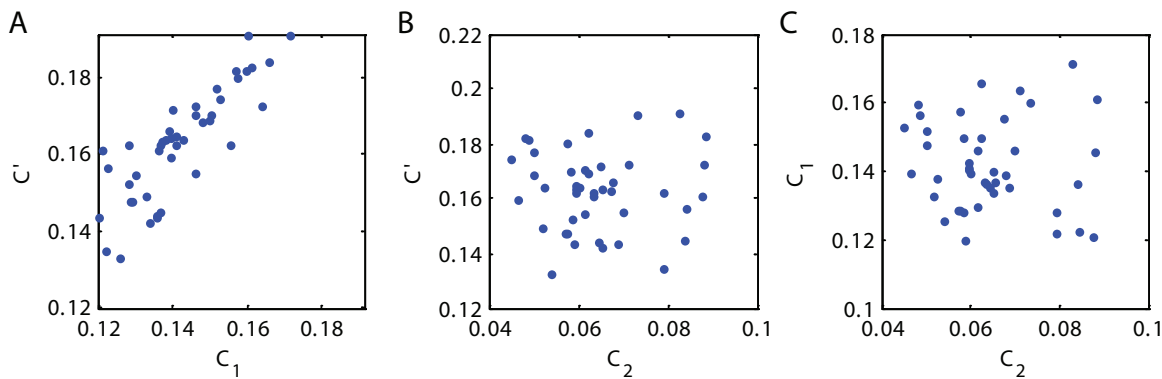}
\caption{Relations between different measures for the clustering coefficient in 43 metabolic networks. Whereas there is a linear correlation ($r = 0.84$) between the new definition $C'$ and $C_1$ (A), there is no correlation between $C'$ and $C_2$ (B, $r = 0.004$) or $C_1$ and $C_2$ (C, $r = -0.06$).}
\label{fig.2}
\end{figure}

We examined the effect of the adjusted definition for the case of comparing sparse networks by analyzing 43 metabolic networks~\cite{Jeong2000}. Testing all 903 distinct relations between pairs of networks, the relations changed -- using the adjusted definition --  in 58\% of the cases for the standard clustering $C_1$. For the alternative more widely used definition $C_2$, the relation changes in 76\% of the cases. Even switching between the traditional definitions $C_1$ and $C_2$ changed the relation in 77\% of the cases.  Comparing the different measures for all 43 networks, there was a linear correlation between $C'$ and $C_1$ but not between $C'$ and $C_2$ or $C_1$ and $C_2$ (Fig.~\ref{fig.2}). This indicates that the effect of using a different clustering coefficient definition can often not be predicted from an existing measure (factors of increase for switching from $C_1$ and $C_2$ to $C'$ is shown in Tab.~\ref{tab.1}).

Another way of comparing neighborhood clustering between networks is the use of clustering coefficient functions. One such clustering coefficient function is $C(k)$ where $k$ is the degree of a node and $C(k)$ is the average clustering coefficient over all nodes with degree $k$ \cite{Chowell2003,Yang2006}. Then, the distributions of $C(k)$ with $k>1$ for the two networks can be compared. Such a comparison might detect cases where one network shows a linear, exponential, or power-law distribution whereas the other network does not. Comparing two networks with a similar distribution becomes more difficult. Whereas qualitative differences might be visible through comparing the distribution plots, getting a quantitative value for describing these differences is more challenging. Therefore, single values for describing networks will remain popular unless standard ways for distribution comparisons are established.

\begin{table}
\caption{Ratios $f$ of the adjusted clustering coefficient $C'$ of 43 metabolic networks with the Watts-Strogatz ($C_1$) and Newman-Strogatz-Watts ($C_2$) clustering coefficient.}
\label{tab.1}
\begin{center}
\begin{tabular}{lllll}
$f$ &		   Mean 	&	   Median  &	  Minimum & Maximum	\\
\hline
$C_1$ & 		1.1510  &  	1.1484  &  	1.0479  & 1.3315\\
$C_2$ & 		2.6241  &  	2.5673  &	  1.7003  & 3.8999\\
\hline
\end{tabular}
\end{center}
\end{table}

\subsection{Changes of small-world features}
Many real-world networks show features of small-world networks \cite{Milgram1967}. In these networks, the characteristic path length $L$ remains comparable with random benchmark networks whereas the average connectivity between neighbors (clustering coefficient) $C$ of a node is much higher than for random networks, that means $L \gtrapprox L_{random}$ but $C\gg C_{random}$ \cite{Watts1998}. One way to assess the extent of small-world features is calculating the small-worldness $s=(C/C_{random}) / (L/L_{random})$ (note that a comparison of small-worldness is only meaningful for similar edge densities as the edge density influences the possible increase in the clustering coefficient). How do these small-world features, in particular the clustering coefficient component of the small-worldness $s$, change with the definition of the clustering coefficient?

Using the measure $C'$ would lead to higher small-worldness $s$--the ratio between the clustering coefficient in the original and a comparable random network--if numerator $C$ for the original networks increases at least as much as the denominator $C_{random}$ for the random benchmark networks. We tested the increase of changing from definition $C_1$ to $C'$ which can be calculated by the ratio $\theta$ (cf. equation \ref{eq.4}): A larger value of $\theta$ results in a larger increase of the clustering coefficient. Therefore, the small-worldness would increase as long as $\theta$ is larger for the original rather than the random benchmark network.

\begin{figure*}
\includegraphics[width=\textwidth]{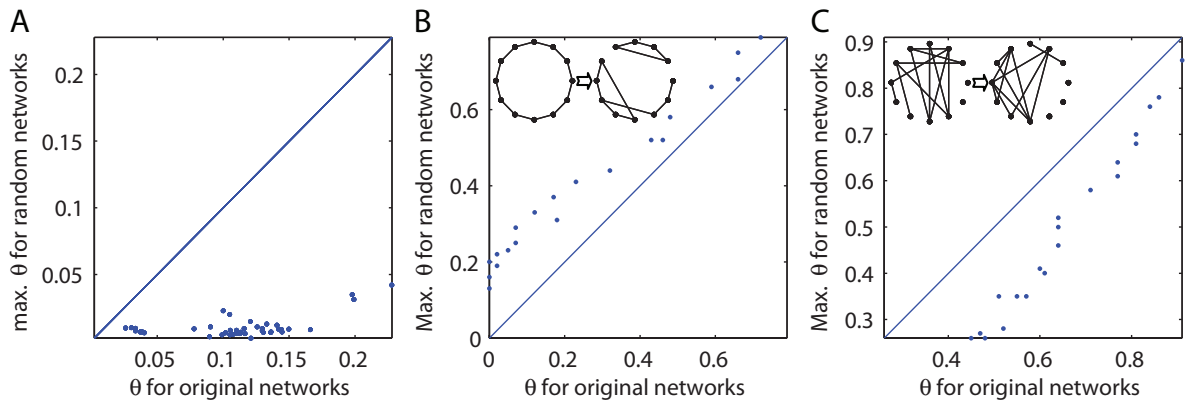}
\caption{Change of small-worldness. Using $C'$ leads to higher small-worldness $s$ if dots are below the identity line and to lower small-worldness above the line. (A) $\theta$ for 43 metabolic networks. (B) $\theta$ for small-world networks generated by rewiring starting with a lattice model \cite{Watts1998} (inset) with 20 different edge densities (maximum $\theta$ of 50 generated networks each). (C) $\theta$ for small-world networks generated by condensation (inverse rewiring) starting with a random model (inset) with 20 different edge densities (maximum $\theta$ of 50 generated networks each).}
\label{fig.4}
\end{figure*}

We tested the increase for the 43 metabolic networks by generating 50 random networks for each metabolic network and using the maximum value of $\theta$ out of the random networks.  For all 43 metabolic networks, $\theta$ was larger for the original network than for random benchmark networks (Fig.~\ref{fig.4}A). 

We also generated artificial small-world networks with 100 nodes and a variable edge density ranging from the minimum (0.5\%) to the maximum (2\%) value of the metabolic networks. For each edge density, 20 small-world networks were generated and for each such small-world network, 50 comparable random networks were analyzed. In contrast to the previous results of real-world networks, random networks show a higher ratio of leafs and isolated nodes than the generated small-world networks (Fig.~\ref{fig.4}B). The reason is that the small-world networks were generated starting from a lattice model followed by random rewiring of the network \cite{Watts1998}. Despite the rewiring, the strong neighborhood connectivity of the lattice model remains and prevents the occurrence of leafs and isolated nodes.

To remove the effect of the lattice network being the starting point of rewiring, we developed a small-world network generator with \textit{inverse rewiring}\footnote{The Matlab script is available at http://www.biological-networks.org/}: the model starts with a random network and rewires edges so that the connectedness but also the number of isolated nodes increases. For a given network with $E$ edges, $10\times E$ rewiring steps were performed. At each step, an existing edge is chosen and deleted. Thereafter, another existing edge is chosen and the starting node of that edge is connected with a randomly chosen node that has not before been connected to that node. Each step elongates an existing chain of nodes by adding an edge, potentially leading to the formation of triangles, whereas the removed edge is either the internal or terminal part of a chain, leading towards a leaf node. For this model, in accordance with the results from the real-world metabolic networks, $\theta$ for the generated small-world network was below the value for random networks (Fig.~\ref{fig.4}C).

\subsection{Changes of small-world classification}
We have seen in the previous section that the small-worldness $s$ of a network increases, or at least stays the same, when the new measure $C'$ is used compared to the classical measure $C_1$. Looking at both measures $C_1$ and $C_2$ could it be the case that networks that were previously classified as random would be classified as small-world with the new measure $C'$? This would be the case if the ratio $C/C_{random}$ is lower or close to 1 for the classical measures but much higher than 1 for $C'$.

\begin{figure}
\includegraphics[width=\textwidth]{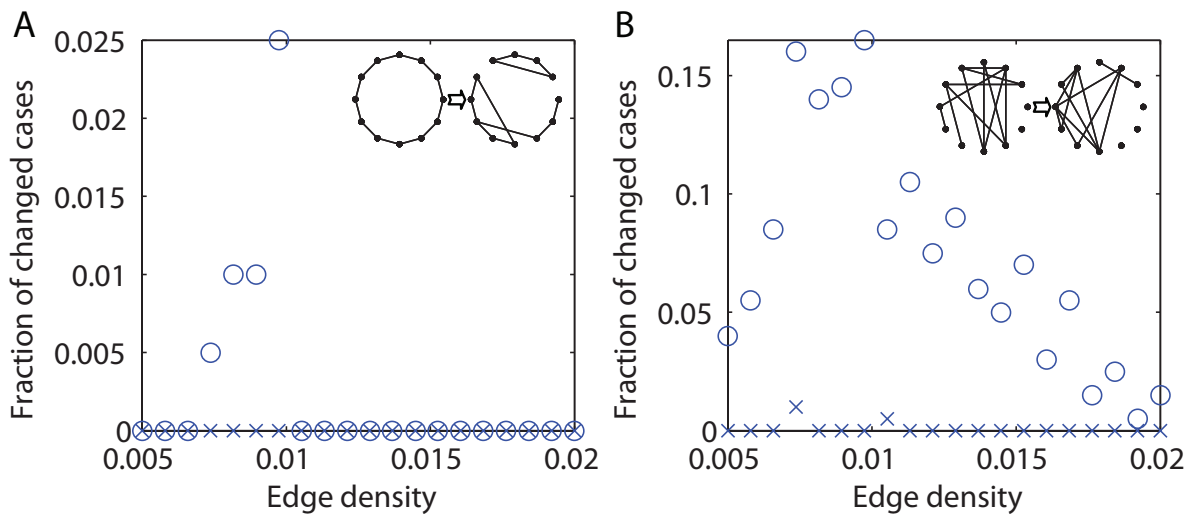}
\caption{Fraction of cases when the network classification changed from random to small-world when switching from $C_1$ (x markers) or $C_2$ (o markers) to $C'$. (A) small-world networks generated by rewiring starting with a lattice model (inset) (B) small-world networks generated by condensation starting with a random model (inset).}
\label{fig.classification}
\end{figure}

Again we tested artificial small-world networks with 100 nodes and a variable edge density using standard and inverse rewiring as described above. For each edge density, 200 networks were generated and 100 benchmark random networks were evaluated for each generated network. The definition of a change in classification from random to small-world was a clustering coefficient ratio $\leq 1$ for $C_1$ or $C_2$ and a ratio $>2$ when using the measure $C'$. For standard rewiring (Fig.~\ref{fig.classification}A), the fraction of changed cases was zero except for a small range of edge densities where the classification changed in up to 2.5\% of the cases when shifting from $C_2$ to $C'$. For inverse rewiring (Fig.~\ref{fig.classification}B), however, classification changed in around 10\% of the cases (up to 15\% for some edge densities) when shifting from $C_2$ to $C'$ in the edge density range of 0.6\%-1.5\%. For shifting from $C_1$ to $C'$, classification only changed up to 1\% of the cases; a fluctuation which might be due to the small sample size. In addition, a shift also occurred between the classical measures $C_1$ and $C_2$: Whereas changing from $C_1$ to $C_2$ affected few cases, a shift from $C_2$ to $C_1$ affected up to 3\% of the cases for standard rewiring and up to 14\% for inverse rewiring. Therefore, changes in classification are possible for all clustering measures; in particular when using inverse rewiring.

\section{Discussion}    
We have shown that current definitions underestimate neighborhood clustering in sparse networks with many isolated or leaf nodes. In addition, the outcome of comparisons of the extent of small-world features between different networks critically depended on the applied definition of the clustering coefficient. Furthermore, networks formerly classified as random can be classified as small-world when isolated or leaf nodes are excluded from the calculation of the average clustering coefficient. This can also happen when switching from $C_2$ to $C_1$.

Could the clustering coefficient definitions impact the analysis of small-world networks?  There are three consequences of this study. First, small-world networks regarding previous measures $C_1$ and $C_2$ will still be detected as small-world using $C'$ as this value will be higher than the previous values. Consequently, the small-worldness $s$---the ratio of the clustering coefficient in the original and random benchmark networks divided by the unchanged ratio of the characteristic path lengths in original and random networks---will be higher. Second, networks which are currently not classified as small-world networks may be regarded as small-world due to the increase in clustering coefficient. This case will occur when the path length is comparable to that of random networks but the clustering coefficient, concerning previous definitions $C_1$ and $C_2$, is not significantly higher than that of random networks. Third, comparison of networks could lead to opposite conclusions using the new measure. In conclusion, the novel measure $C'$ gives a clearer view of neighborhood connectivity and is more independent of the sparseness of edge density. 

A problem of the proposed measure $C'$is that the percentage $\theta$ of nodes that are excluded from analysis could be considerably high (Tab.~\ref{tab.2}). The percentage of excluded nodes could be as high as 14\% for metabolic networks and as high as 59\% for man-made networks (power grid). Note that the value for the protein interaction network in yeast is also high at 56\% as edge density is low and isolated nodes are not part of the largest connected cluster observed here. In general, however, exclusion from the average affected less than 10\% for most of the networks. 
In addition, using a subset of defined nodes is comparable to the procedure for calculating shortest paths or the characteristic path length where unreachable paths with otherwise infinite distance are not included in calculating the average path length.  

\begin{figure}
\includegraphics{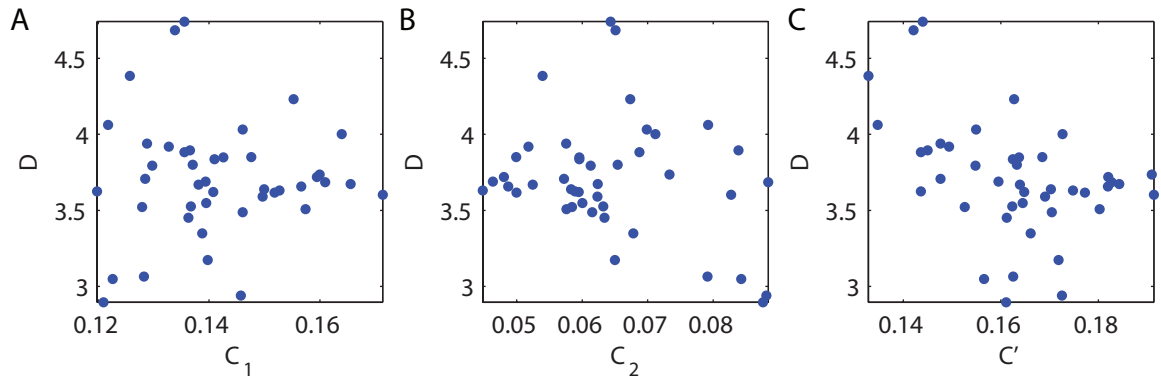}
\caption{Relations between disconnectedness $D$ and measures for the clustering coefficient in 43 metabolic networks. Correlations are $r = 0.04$ between $C_1$ and $D$ (A),  $r = -0.28$ between $C_2$ and $D$ (B), and $r = -0.43$ between $C'$ and $D$ (C).}
\label{fig.5}
\end{figure}

An alternative solution would be to describe the clustering coefficient using inverse neighborhood clustering. For the shortest paths, for example, the inverse measure of efficiency \cite{Achard2007} where unreachable paths contribute $1/\infty = 0$ to the local efficiency circumvents the need for excluding unreachable paths. Similarly, the (neighborhood) disconnectedness $D$ could be defined as:
\begin{equation}
\label{eq.5}
D=\frac{1}{N}\sum D_i \quad\mbox{with}\quad D_i=1/C_i=\frac{deg_i(deg_i-1)}{\Gamma_i}
\end{equation}
$$ \quad\mbox{and}\quad D_i=0\quad\mbox{for}\quad \Gamma_i=0 $$
Here, nodes which are leafs or isolated will contribute a zero value to the average $D$ as one of the degrees will be zero for these nodes. $D$ will be high when neighbors are not connected and low ($\to$1) for high connectivity between neighbors. The correlations between disconnectedness and measures of the clustering coefficient are shown in Figure~\ref{fig.5}.

\section{Conclusion} 
Including the percentage $\theta$ in publications could help to understand the validity of the applied definition of the clustering coefficient regardless of whether it is the Watts-Strogatz definition $C_1$, the Newman-Strogatz-Watts definition $C_2$, or the alternative definition $C'$ presented here. In addition, this information is critical for the classification as small-world networks. We therefore suggest that information about the applied definition and the number of leafs and isolated nodes, the ratio $\theta$, should be included in addition to the value of the average clustering coefficient.

\ack 
Supported by EPSRC (EP/E002331/1) and Royal Society (RG/2006/R2). \\

\bibliographystyle{unsrt}

\newcommand{\noopsort}[1]{} \newcommand{\printfirst}[2]{#1}
  \newcommand{\singleletter}[1]{#1} \newcommand{\switchargs}[2]{#2#1}

\end{document}